
\magnification=1200
\hsize=6truein
\vsize=9.truein
\hoffset=.5truein
\voffset=.5truecm
\baselineskip=.7truecm
\nopagenumbers
\def\Dir{\kern -6.4pt\Big{/}}
\def\DDir{\kern -7.6pt\Big{/}}
\def\DGir{\kern -6.0pt\Big{/}}
\rightline{ DFTT 53/92\ \ \ }
\smallskip
\rightline{\rm October 1992\ \ \ }
\vskip 30pt
\centerline{\bf HEAVY QUARKS AND LEPTONS AT $e^+e^-$ COLLIDERS}
\vskip .6in
\centerline{\rm Alessandro BALLESTRERO}
\medskip
\centerline{\it INFN, Sezione di Torino, Italy}
\vskip .3in
\centerline{\rm Ezio MAINA}
\medskip
\centerline{\it Dipartimento di Fisica Teorica,
Universit\`a di Torino, Italy}
\centerline{\it and INFN, Sezione di Torino, Italy}
\vskip .3in
\centerline{\rm Stefano MORETTI}
\medskip
\centerline{\it INFN, Sezione di Torino, Italy}
\vskip .45in
\centerline{ ABSTRACT}
\medskip\noindent
The production of massive quarks and leptons in $e^+e^-$ collisions
is studied
using exact helicity amplitudes. Total cross sections as a function of
$y_{\rm cut}$, in both the JADE and the $k_T$ algorithms, are presented
and compared with massless results.
Some invariant mass distributions are examined in relation to Higgs
detection.
Compact expressions for the helicity amplitudes are given.
\footnote{ }{ Work supported in part by Ministero dell' Universit\`a
e della Ricerca Scientifica.}
\par\vfill\eject
\footline={\hss\tenrm\folio\hss}
\par
\centerline{\bf 1. Introduction}
\bigskip
Massive particles are abuntantly produced at $e^+e^-$ colliders.
Often they are associated with other, massless,
particles in rather complicated
final states. Whether masses are important, or whether they can
be ignored, depends from the center of mass energy and also from the
region in phase space which is under study.
For instance, the electron mass can be ignored in
most cases at LEP energies, but for its role in regulating collinear
divergencies in reactions like Bhabha scattering at small angle.
On the contrary, the top mass is so large that it must be included in
all cases even at supercollider energies.
In a recent paper [1] we have shown that at the $Z^0$ peak, when
masses are properly taken into account, cross sections involving
$b$ or, to a lesser extent, $c$ quarks differ significantly from
the corresponding predictions obtained when masses
are neglected.
It is the purpose of this paper
to extend these results in several directions.
First of all we consider both LEP I and LEP II energies. At higher
center of mass energy, mass effects are expected to be smaller,
but only an explicit calculation can establish to which degree
this is correct. On the
other hand, one of the main task of LEP II will be to search for the
Higgs boson in the mass range
45 GeV $\leq \hskip 2pt m_H \hskip 2pt\leq$ 80 GeV [2].
Such a Higgs decays almost exclusively to $b\bar b$ and $\tau^+ \tau^-$
and as a consequence it is important to determine as accurately as
possible all reactions in which $b\bar b$ and $\tau^+ \tau^-$ pairs
are produced, which could provide a background to Higgs detection.
Second, we extend the study of the production of up to four
strongly interacting particles. Four--quark
final states with quarks of different flavours are also examined.
A number of jet--jet mass distributions are given,
and the effects of different clustering algorithms
are investigated.
Third, we consider the
production of two quarks and two leptons and of four leptons.
This kind of events are relevant as a potential background to
Higgs searches and provide a test of the Standard Model, though
not at the level of precision which can be reached in more inclusive
measurements.
Fourth, we consider final states including photons which are
actively studied experimentally in order to determine the electroweak
couplings of the quarks, and as a mean to search for new
phenomena.
Finally we present the helicity amplitudes for all the reactions
studied in this paper.
\par
We only consider processes without electrons in the final state,
which proceed through $e^+e^-$ annihilation to photon or $Z^0$.
The two contribution are comparable outside the
$Z^0$ peak. Therefore both have been retained in all amplitudes.
\par
The powerful methods based on helicity amplitudes, which have
enormously simplified the calculation of processes with several
final state particles, can, in principle, handle massless and massive
fermions with equal ease. In practice, however, the massless case
has proven to be much simpler and a number of very elegant and
physically relevant results have been obtained in this case.
For a good introduction and complete references, see [3].
The treatment of reactions involving massive fermions is still
much less advanced.\par
Several papers have studied three and four particle production at $e^+e^-$
colliders [5-11] at the matrix element level but, to our knowledge,
all of them lacked an essential ingredient, either considering
only massless particles or neglecting the $Z^0$ contribution,
for a consistent treatment of heavy particle production at LEP
energies.\par
The matrix element for all processes for which we give results
has been computed at tree level
following both the method of ref.[12,13] and that of ref.[14]
as a check of the correctness of our results.
In Appendix B we present the amplitudes in the formalism
of ref.[12,13] which makes it simpler to give
compact, easy to implement formulae.
In Appendix A, for the convenience of the reader,
we briefly recall the method and the results
of ref.[12,13] and we collect various formulae which are used
in the analytic expressions.
\par
The amplitudes have been checked for gauge invariance, and
for BRST invariance [15] in case of external $Z^0$'s.
In the appropriate limits our results reproduce
those of ref.[4,9,11,26] after some misprints in the formulae of
ref.[11] have been corrected.\par
We have used  $M_Z=91.1$ GeV, $\Gamma_Z=2.5$ GeV,
$\sin^2 (\theta_W)=.23$, $\alpha_s= .115$, $\alpha_{em}= 1/128$,
$m_\tau=1.78$ GeV, $m_c=1.7$ GeV and $m_b=5.$ GeV
in the numerical part of our work.\par
In what follows we neglect all hadronization effects, and
apply cuts at the partonic level.\par
\bigskip
\centerline{\bf 2. Jets }
\bigskip
All LEP experiments have performed a large number of QCD tests.
For instance, mentioning only the measurements which are most likely
to be sensitive to masses, $\alpha_s$ has been determined from jet
rates [16] and flavour independence of the coupling has been verified [17].
Three [18] and four [19] jet distributions have been studied and compared
with QCD predictions. The color factors, which determine the gauge
group which is responsible for strong interactions, have been measured
[20].
The possibility of tagging quark jets using the semileptonic
decays of $b$ and $c$ quarks has been exploited, for example in studies of
the differences between gluon and quark jets [21]. \par\noindent


In the coming years, improvements in statistics, in secondary vertex
reconstruction with silicon vertex detectors and in particle
identification will allow much more detailed studies of heavy quark
production at LEP.\par
The experimental definition of a jet is based on a clustering procedure.
The two most widely used schemes are the JADE algorithm [22] based on the
variable
\smallskip
$$ y^J_{ij} = 2 {{E_i E_j}\over{E^2_{\rm vis}}} ( 1 - \cos\theta_{ij} )
\eqno(1)$$
\smallskip\noindent
and the $k_T$ or Durham algorithm [23] which makes use of
a new clustering variable
\smallskip
$$ y^T_{ij} = 2 {{\min (E^2_i, E^2_j)}
\over{E^2_{\rm vis}}} ( 1 - \cos\theta_{ij} ). \eqno(2)$$
\smallskip\noindent
In fig.1 we present the cross sections for $e^+e^-\rightarrow q \bar q g$
and $e^+e^-\rightarrow q \bar q gg$ with
$q =  d, b$ as a function of $y_{\rm cut}$ for both definitions of $y$
at LEP I.
For small $y^T_{\rm cut}$ the cross
section for $b\bar b g$ is almost 20\% smaller than for $d \bar d g$.
As expected the ratio becomes closer to one for larger $y_{\rm cut}$,
but for $y_{\rm cut}$ as large as .2,
still $R^{bd}_3 = \sigma (b \bar b  g)/\sigma (d \bar d g )\leq .96$
in both schemes.
The cross sections for the same reactions at $\sqrt s = 200$ GeV
are given in fig.2. The differences between the massive and the massless
case are less important at higher energies, as expected, but still
of the order of several percent. It has to be noticed that the Durham scheme
tends to enhance these differences. This feature has been found in all
cross sections and distributions we have studied. Therefore,
depending on the kind of analysis which is performed on the data sample,
different schemes can be used in order to exalt or suppress
mass effects.
A typical example of this behaviour
is shown in fig.3a and 3b where the invariant mass
distributions in three jet events are presented. The different shape of
of $b$ and $d$ events is quite noticeable for the Durham algorithm,
not only for the invariant mass of the $qg$ pair, which is expected
because of the different quark energy threshold, but also for the
invariant mass of the $q\bar q$ pair. The exact shape of these
distributions depends on $y_{\rm cut}$ but the different sensitivity
to mass effects in the two schemes remains.\par
The $q \bar q gg$ final state
is the dominant contribution to the four--jet cross section.
The process with two hard gluon emissions is more sensitive to masses
than the process with only one emission. For instance
 $R^{bd}_4 = \sigma (b \bar b g g)/\sigma (d \bar d gg ) =  .7$
for $y^T_{\rm cut} = 1.5 \times 10^{-3}$.
\par
The various contributions to $e^+e^-\rightarrow q \bar q q' \bar q' $
are shown in fig.4 ($q = q'$) and fig.5 ($q \ne q'$) as a function
of $y_{\rm cut}$.
Though very small compared with the $q \bar q gg$ rates these
cross sections could be
quite interesting if heavy quarks can be tagged with high efficiency.
For $10^6$ $Z^0$, corresponding to an integrated luminosity of about
30 $pb^{-1}$, one expects approximately 100 events with 4$b$ quarks
and 400 events with 2$b$ and 2$c$ quarks at $y_{\rm cut}^J = 1.\times
10^{-2}$.
The differences between massive and massless results are large, and even the
charm mass has a significant effect. It is interesting to notice that the
cross section for $b \bar b u \bar u$ is larger than the
cross section for $b \bar b d \bar d$, contrary to the naive
expectations based on the fact that down-type quarks couple more strongly
to the $Z^0$ than up-type quarks. This is due to the interference
between graphs 1 and 2 with graphs 3 and 4 of fig. 12a.
\par
{}From fig.1, 2, 4 and 5 it can be seen that, while for  three parton
processes the cross
sections for $y^J_{\rm cut} = 1. \times 10^{-2}$ are approximately
matched by the cross sections
for $y^T_{\rm cut} = 1.5 \times 10^{-3}$, in the four parton case this happens
for $y^T_{\rm cut} = 3. \times 10^{-3}$.\par
The analytic expression of the matrix element for $e^+e^-\rightarrow q \bar q
g$
obtained in ref.[4], neglecting the $Z^0$, has been exploited by some
experimental
groups [17] to estimate mass corrections to jet rate predictions which are
used for
measuring $\alpha_s$. The theoretical three jet rate to order $O(\alpha^2_s)$
for $d$ quarks is multiplied by the ratio of the
$b \bar b g$ to $d \bar d g$ tree level cross sections, $R^{bd}_3$
after full detector simulation. We have
compared the ratio obtained taking into account all contributions with the
ratio
obtained neglecting the $Z^0$. The latter is consistently about 1\% lower than
the former at all $y_{\rm cut}$. A further comment is in order. The theoretical
three jet rate is a combination of three and four parton processes,
the four parton contribution being of the order of 20\% in the Durham scheme
at small $y_{\rm cut}$.
{}From our figures it is clear than the mass correction factors for three and
four
parton reactions differ considerably and therefore the simple method
adopted in order to evaluate mass corrections might slightly
underestimate them. \par
\bigskip
\centerline{\bf 3. Backgrounds to Higgs searches }
\bigskip
The main production mechanism [2] for the Higgs particle
at energies between 170 and 200 GeV
is $e^+e^-\rightarrow H Z$. If the Higgs mass is between 40
and 80 GeV the cross section for this process is of the order of 1 $pb$.
As already mentioned the Higgs decays predominantly to $b \bar b$.
For an integrated luminosity of 500 $pb^{-1}$ at $\sqrt s$ = 200 GeV
and $m_H$ = 60 GeV this yields about 30 events for
$e^+e^-\rightarrow H Z\rightarrow b \bar b l^+ l^-$ ($l = e, \mu$),
80 events for
$e^+e^-\rightarrow H Z\rightarrow b \bar b \nu \bar \nu$
and 340 events for
$e^+e^-\rightarrow H Z\rightarrow b \bar b j j$.
Four--jet events are a potential source of background to this last
type of processes. If $b$'s can be efficiently tagged, the main
background contribution is given by
$e^+e^-\rightarrow b \bar b gg$. For this reason we have computed
the invariant mass distributions of all particle pairs in
the $b \bar b gg$ final state. These distributions are shown
in figs.6, 7 and 8
at $\sqrt s$ = 91.1, 170 and 200 GeV respectively, in the two schemes.
The intermediate point has been studied as a realistic energy for the
first phase of LEP II.
We have required $y^J_{ij}\geq 1.\times 10^{-2}$ or
$y^T_{ij}\geq 1.5\times 10^{-3}$ for all pairs
and that the angle of each particle with the beam satisfies
$\mid \cos( \theta ) \mid < .95$.
Again one notices the differences between the two schemes, particularly for
small invariant masses.
At LEP II energies the $b \bar b$ mass peaks well above 100 GeV
while the $gg$ mass clusters at small invariant masses.
The cross section at $\sqrt s = 200$ GeV,
for 40 GeV $\leq m_{b \bar b} \leq$ 80 GeV
is only about $5\times 10^{-2}\hskip 2pt pb$
which corresponds to approximately 25 events.
Moreover since the two gluons should fake an hadronic decay of the
$Z^0$, one could impose a cut on their invariant mass,
decreasing this background drastically.
\par
The Higgs production cross section times the branching ratio to $\tau^+
\tau^-$ at LEP II energies
is about 80 $fb$. Hence it might be possible to detect the
Higgs in this channel and to measure its coupling to the
$\tau$. A comparison of this decay mode with the dominant one to $b \bar b$
would test the predicted proportionality of Higgs coupling to fermion
masses.
In fig.9 we present the invariant mass distribution of the $\tau^+
\tau^-$ pair in $e^+e^-\rightarrow \tau^+\tau^- q\bar q$ at $\sqrt s =
170$ GeV summed over five massless flavours. In order to select events which
could fake an hadronic decay of the $Z^0$
we have required the invariant mass of the $q \bar q$
pair to be larger than 60 GeV.
The cross section
integrated for $\tau^+ \tau^-$ masses between 40 and 70 GeV is less
than 10 $fb$. Hence this background is rather small.
On the other hand, with the mentioned cut,
$\sigma(e^+e^-\rightarrow \tau^+\tau^- q\bar q) = 86 \hskip 2 pt fb$.
This result is comparable to the Higgs production
cross section in the $\tau^+\tau^-$ channel and it can be used as
a reference point for the Higgs search.
\par
\bigskip
\centerline{\bf 4. Jets plus Leptons and Four Leptons }
\bigskip
Four--fermion final states with at least one lepton pair are relatively easy to
study experimentally, and in some cases they represent a
potential background to Higgs production.
Much interest has been spurred by the observation of an
apparent excess in the $\tau^+\tau^-X$ channel in ALEPH's 1989-1990 data [24].
This observation has not been confirmed by other collaborations [25].
\par
Quite recently a new Montecarlo [26], which includes fermion masses
and a complete treatment of $\gamma$ and $Z^0$ contributions
as is appropriate outside the $Z^0$ peak,
and combines them with initial and final state
radiation and with a careful mapping
of the many peaks in the matrix element due to collinear configurations,
has been presented. We have computed the cross section for
$e^+e^-\rightarrow \mu^+\mu^-\tau^+\tau^- $ and
$e^+e^-\rightarrow \tau^+\tau^-\tau^+\tau^- $, without initial
state radiation,
with the parameters used in ref.[26] and we obtain
$\sigma_{tot}(e^+e^-\rightarrow \mu^+\mu^-\tau^+\tau^-) = 640.
\pm 7. \hskip 2pt fb$
and
$\sigma_{tot}(e^+e^-\rightarrow \tau^+\tau^-\tau^+\tau^-) = 61.2
\pm .3 \hskip 2pt fb$,
in excellent agreement with that reference. Since we are primarily interested
in massive particle production we have made no special effort to
describe collinear configurations and have used the standard VEGAS [27]
integration routine with the same parametrization over the whole phase
space. Unfortunately our straightforward approach does not allow
us to obtain numerically stable results for
$e^+e^-\rightarrow \mu^+\mu^-\mu^+\mu^- $ with the cuts of ref.[26].
Moreover we have not produced the additional diagrams
needed to describe
four--fermion production with final state electrons nor
we have implemented initial state radiation, and this
prevents us from a more thorough
comparison between our results and those of ref.[26].\par
In table I we give the total cross sections for
$e^+e^-\rightarrow \tau^+\tau^-\tau^+\tau^-$,
$e^+e^-\rightarrow \tau^+\tau^- c\bar c$
and $e^+e^-\rightarrow \tau^+\tau^- b\bar b$
at $\sqrt s = 91.1$ and 200 GeV with the standard parameters
used in this paper, in particular with $\alpha_{em}= 1/128$
at all electroweak vertices. The expected number of events is very small,
between 1 and 5 events per $10^6$ $Z^0$ at LEP I.\par
\vfill\eject
\centerline{\bf 5. Final states including a Photon or $Z^0$ }
\bigskip
Photons are the only particles which can be directly revealed
and couple to the quarks in the
early stages of the hadronization process.
Final state radiation in multihadronic decays of the $Z^0$ can
be used to test the electroweak
couplings of up and down type quarks [28,29],
combining the measurement of the radiation rate with the
measured hadronic width of the $Z^0$. \par
In fig.10 we give the cross sections for $e^+e^-\rightarrow q \bar q \gamma$
and $e^+e^-\rightarrow q \bar q \gamma g$, with $q = d, b$, at the $Z^0$ peak
as a function of $y_{\rm cut}$.
The photon transverse momentum is required to be greater than 5 GeV.
The difference between $b$ and $d$ quarks is again quite important,
particularly for small $y_{\rm cut}$.  These differences are also visible in
the
photon energy spectrum given in fig.11a and 11b for the JADE and Durham scheme,
respectively, even though the overall shape of the distributions are similar.
\par
In table II we present the total cross section for production
of a $Z^0$ in association with a heavy fermion pair at
$\sqrt s = 200$ GeV. We implicitly assume that the $Z^0$ decays to a
fermion pair different from the one that appears in the event.
These processes are dominated by the diagrams in fig.13b.
Our results show that both intermediate photon and $Z^0$ give a
sizable contribution. In fact if only diagrams with two $Z^0$'s
were important one would expect $\sigma (e^+e^-\rightarrow Z \mu^+ \mu^-)
\approx \sigma (e^+e^-\rightarrow Z \tau^+ \tau^-)$;
if, on the contrary, only photonic intermediate states were
relevant one would get $\sigma (e^+e^-\rightarrow Z c \bar c)
\approx 3\times (2/3)^2 \sigma (e^+e^-\rightarrow Z \tau^+ \tau^-)$.
All processes produce rates of the order of a hundred events for
500 $pb^{-1}$ of integrated luminosity.
\bigskip
\centerline{\bf 6. Conclusions}
\bigskip
We have computed the exact matrix elements, at tree level,
for all processes at $e^+e^-$ colliders with three and four
particles in the final state in which masses are relevant,
with the only exception of reactions with final state electrons.
We have studied these processes at LEP I and LEP II energies
using both the JADE and the Durham clustering algorithm.\par
Total cross sections involving $b$ quarks are substantially smaller
than the corresponding ones for $d$ quarks, particularly for
small $y_{\rm cut}$. The effect increases with the number
of jets and (obviously) with the number of massive particles.
The differences between the cross sections or distributions
for massive quarks and those with massless ones are generally larger
in the Durham algorithm.\par
Four--jet events involving heavy particles and two--tau--two--jet events
are a possible background to Higgs production. Therefore it is important to
have exact expressions for all these reactions.
We have studied several mass distributions in connection
with the discovery of the Higgs and the measurement of its couplings.
In the relevant mass regions these processes
have rates much smaller than Higgs production.\par
We have presented cross sections and energy distributions for
one photon production in association with up to three jets,
keeping quark masses into account. As in the case of purely hadronic
final states, the overall shape of distributions is not strongly modified by
mass effects, which however yield substantially lower rates than
the corresponding reactions for massless particles.\par
It will soon be possible, with improved statistics and tagging
capabilities, to put to the test most of these predictions.
\par\vfill\eject
\vskip 30pt

\centerline{\bf Appendix A: The helicity amplitudes method}

\vskip.25in

In this section, for completeness, we briefly recall the spinor techniques
of ref.[12,13] which are used in our calculations.

\vskip.15in

{  (i)} {\it Spinors.}
External fermions$^{(1)}$ of mass $m$ and momentum $p^\mu$
are described by spinors
corresponding to states of definite helicity $\lambda$,
$u(p,\lambda)$ verifying the Dirac equation

\footnote{}{(1) Unless stated otherwise, we shall use the term
``fermion'' and the symbol $u$ for both particles and antiparticles.}

$$\eqalign{p\Dir u(p,\lambda)&=\pm m u(p,\lambda),\cr
\bar u(p,\lambda)p\Dir &=\pm m \bar u(p,\lambda),\cr}\eqno(A.1)$$

and the spin sum relation

$$\sum_{\lambda=\pm} u(p,\lambda)\bar u(p,\lambda)=
p\Dir\pm m,\eqno(A.2)$$

where the sign $+(-)$ refers to a particle (antiparticle).

One can choose two arbitrary
vectors $k_0$  and $k_1$ such that

$$ k_0\cdot k_0=0, \quad\quad k_1\cdot k_1=-1, \quad\quad k_0\cdot k_1=0,
\eqno(A.3)$$

and express the
spinors $u(p,\lambda)$ in terms of chiral ones
$w(k_0,\lambda)$ as

$$u(p,\lambda)=w(p,\lambda)+\mu w(k_0,-\lambda),\eqno(A.4)$$

where

$$w(p,\lambda)=p\Dir w(k_0,-\lambda)/\eta,\eqno(A.5)$$

and

$$ \mu=\pm {m\over{\eta}}, \quad\quad\quad \eta=\sqrt{2(p\cdot k_0)}.
\eqno(A.6)$$

The spinors $w(k_0,\lambda)$ satisfy

$$w(k_0,\lambda)\bar w(k_0,\lambda)=
{{1+\lambda\gamma_5}\over{2}}{k\Dir}_0,\eqno(A.7)$$

and therefore

$$\sum_{\lambda=\pm}w(k_0,\lambda)\bar w(k_0,\lambda)=
{k\Dir}_0.\eqno(A.8)$$

The phase between chiral states is fixed by

$$w(k_0,\lambda)=\lambda {k\Dir}_1 w(k_0,-\lambda).\eqno(A.9)$$

The freedom in choosing $k_0$ and $k_1$ provides a powerful tool
for checking the correctness of any calculation.

\vskip.15in

{(ii)} {\it Polarization vectors for massless gauge bosons.}
External spin 1 massless gauge bosons of momentum $p^\mu$ are described
by polarization vectors
corresponding to states of definite helicity $\lambda$,
$\varepsilon^{\mu}(p,\lambda)$ satisfying

$$\varepsilon(p,\lambda)\cdot p=0,
\quad\quad\quad
\varepsilon(p,\lambda)\cdot \varepsilon(p,\lambda)=0,$$

$$\varepsilon^\mu(p,-\lambda)=\varepsilon^{\mu *}(p,\lambda),
\quad\quad\quad
\varepsilon(p,\lambda)\cdot \varepsilon(p,-\lambda)=-1,\eqno(A.10)$$

and the spin sum relation (in the axial gauge)

$$\sum_{\lambda=\pm}
\varepsilon^\mu(p,\lambda)
\varepsilon^{\nu *}(p,\lambda)=-g^{\mu\nu}+
{{q^\mu p^\nu+q^\nu p^\mu}\over{p\cdot q}},\eqno(A.11)$$

where $q^\mu$ is any four--vector not proportional to $p^{\mu}$.

Any objects
$\varepsilon^\mu(p,\lambda)$ obeying the relations (A.10)--(A.11)
make an acceptable choice for the polarization vectors. For instance

$$\varepsilon^\mu(p,\lambda)=N
[\bar u(q,\lambda)\gamma^\mu u(p,\lambda)],\eqno(A.12)$$

$N$ being the normalization factor

$$N=[4(q\cdot p)]^{-1/2}.\eqno(A.13)$$

The existing freedom in choosing $q^\mu$ corresponds to fixing the gauge.
The final results do not depend on the choice of $q^\mu$.

\vskip.15in

{ (iii)} {\it Polarization vectors for massive gauge bosons.}
For spin 1 massive gauge bosons
there is an additional longitudinally polarized state satisfying

$$\varepsilon(p,0)\cdot p=0,
\quad\quad
\varepsilon(p,\lambda)\cdot \varepsilon(p,0)=0,
\quad\quad
\varepsilon(p,0)\cdot \varepsilon(p,0)=-1.\eqno(A.14)$$

The spin sum becomes

$$\sum_{\lambda=\pm,0}
\varepsilon^\mu(p,\lambda)
\varepsilon^{\nu *}(p,\lambda)=-g^{\mu\nu}+
{{p^\mu p^\nu}\over{m^2}},\eqno(A.15)$$

where $m$ and $p^\mu$ are the gauge boson mass and
momentum, respectively.

For the
polarization vectors of
massive gauge bosons we cannot adopt the form (A.12), since
for a timelike momentum $p^\mu$ we cannot assign a
definite helicity in a covariant way to the
spinors $u(p)$ and antispinors $v(p)$.

The simplest solution lies in noting that we are usually dealing with
cross sections for unpolarized bosons, so we really have as the only
requirement on the $\varepsilon^\mu(p,\lambda)$ that their spin sum
should be as in eq.(A.15).
Any way by which we arrive at that expression gives us an acceptable
choice for the polarization representation.

Introducing the quantity

$$a^\mu=\bar u(r_2,-)\gamma^\mu u(r_1,-),\eqno(A.16)$$

where $r_1^\mu$, $r_2^\mu$ are two lightlike four--vectors satisfying

$$r_1^2=r_2^2=0,\quad\quad\quad r_1^\mu+r_2^\mu=p^\mu,\eqno(A.17)$$

this can be used, after proper normalization, as the polarization
vector belonging to $p^\mu$.
In fact, if we replace the spin sum by an integral over the solid angle
$d\Omega$  of $r_1$ in the rest frame of $p^\mu$,
the result is of the desired form:

$$ \int d\Omega~a^\mu a^{\nu *} = {8\over{3}}\pi m^2
(-g^{\mu\nu}+{{p^\mu p^\nu}\over{m^2}}).\eqno(A.18)$$

This implies that we will obtain the correct result for the cross
sections if we make the following replacements for on--shell
massive bosons

$$\varepsilon^\mu \rightarrow a^\mu,\quad\quad\quad
\sum \varepsilon^\mu \varepsilon{^{\nu *}}\rightarrow
{3\over{8\pi m^2}} \int d\Omega~a^\mu a^{\nu *}.\eqno(A.19)$$

\vskip.15in

{ (iv)} {\it $S$ and $Z$ functions.}
Using the previous definitions one can compute

$$S(\lambda,p_1,p_2)=[\bar u(p_1,\lambda) u(p_2,-\lambda)],
\eqno(A.20)$$

and

$$\hskip -1.2in
Z(p_1,\lambda_1;p_2,\lambda_2;p_3,\lambda_3;p_4,\lambda_4;
c_R,c_L;c'_R,c'_L)=$$
$$\hskip1.7in
[\bar u(p_1,\lambda_1) \Gamma^{\mu} u(p_2,\lambda_2)]
[\bar u(p_3,\lambda_3) \Gamma'_{\mu} u(p_4,\lambda_4)],\eqno(A.21)$$

where

$$\Gamma^{(')\mu}=\gamma^{\mu}\Gamma^{(')},\eqno(A.22)$$

and

$$\Gamma^{(')}=(c^{(')}_R P_R + c^{(')}_L P_L),\eqno(A.23)$$

with

$$P_R={{1+\gamma_5}\over{2}},\quad\quad\quad
P_L={{1-\gamma_5}\over{2}},\eqno(A.24)$$

the chiral projectors.

\vskip.15in

The results ($\epsilon^{0123} = 1$) are (see (A.6))
$$S(+,p_1,p_2)= 2{{(p_1\cdot k_0)(p_2\cdot k_1)
 -(p_1\cdot k_1)(p_2\cdot k_0)
 +i\epsilon_{\mu\nu\rho\sigma}
  k^\mu_0k^\nu_1p^\rho_1p^\sigma_2}\over{\eta_1\eta_2}},
\eqno(A.25)$$

$$S(-,p_1,p_2)= S(+,p_2,p_1)^*,
\eqno(A.26)$$

and

$$\hskip-2.in
Z(p_1,+;p_2,+;p_3,+;p_4,+;c_R,c_L;c'_R,c'_L)=$$
$$\hskip.6in
-2[S(+,p_3,p_1)S(-,p_4,p_2)c'_Rc_R
-\mu_1\mu_2\eta_3\eta_4c'_Rc_L
-\eta_1\eta_2\mu_3\mu_4c'_Lc_R],$$

$$\hskip-2.in
Z(p_1,+;p_2,+;p_3,+;p_4,-;c_R,c_L;c'_R,c'_L)=$$
$$\hskip1.in
-2\eta_2c_R[S(+,p_4,p_1)\mu_3c'_L-S(+,p_3,p_1)\mu_4c'_R],$$

$$\hskip-2.in
Z(p_1,+;p_2,+;p_3,-;p_4,+;c_R,c_L;c'_R,c'_L)=$$
$$\hskip1.in
-2\eta_1c_R[S(-,p_2,p_3)\mu_4c'_L-S(-,p_2,p_4)\mu_3c'_R],$$

$$\hskip-2.in
Z(p_1,+;p_2,+;p_3,-;p_4,-;c_R,c_L;c'_R,c'_L)=$$
$$\hskip.6in
-2[S(+,p_1,p_4)S(-,p_2,p_3)c'_Lc_R
-\mu_1\mu_2\eta_3\eta_4c'_Lc_L
-\eta_1\eta_2\mu_3\mu_4c'_Rc_R],$$

$$\hskip-2.in
Z(p_1,+;p_2,-;p_3,+;p_4,+;c_R,c_L;c'_R,c'_L)=$$
$$\hskip.9in
-2\eta_4c'_R[S(+,p_3,p_1)\mu_2c_R-S(+,p_3,p_2)\mu_1c_L],\eqno(A.27)$$

$$\hskip-1.85in
Z(p_1,+;p_2,-;p_3,+;p_4,-;c_R,c_L;c'_R,c'_L)=0,$$

$$\hskip-2.in
Z(p_1,+;p_2,-;p_3,-;p_4,+;c_R,c_L;c'_R,c'_L)=$$
$$\hskip.6in
-2[\mu_1\mu_4\eta_2\eta_3c'_Lc_L
+\mu_2\mu_3\eta_1\eta_4c'_Rc_R
-\mu_2\mu_4\eta_1\eta_3c'_Lc_R
-\mu_1\mu_3\eta_2\eta_4c'_Rc_L],$$

$$\hskip-2.in
Z(p_1,+;p_2,-;p_3,-;p_4,-;c_R,c_L;c'_R,c'_L)=$$
$$\hskip1.in
-2\eta_3c'_L[S(+,p_2,p_4)\mu_1c_L-S(+,p_1,p_4)\mu_2c_R].$$

\vskip.15in

The remaining $Z$ functions can be obtained by exchanging
$+\leftrightarrow -$ and $R\leftrightarrow L$.

\vfill\eject
\centerline{\bf Appendix B: Matrix elements}

\vskip.25in
In this section we give the analytic formulae
for the matrix elements.
We define the propagator functions as
$$D_{\gamma ,g}(p)={1\over {p^2}},
\quad\quad
D_{Z}(p)={1\over
{sin^{2}\theta_W cos^{2}\theta_W(p^2-M_Z^2+iM_Z\Gamma_Z)}},
\eqno(B.1)$$

and

$$D_{f}(p)={1\over {p^2-m_f^2}},\eqno(B.2)$$

where
$M_Z$ and $\Gamma_Z$ are, respectively, the mass and the width
of the $Z$ boson, $\theta_W$ is the weak mixing angle
and $m_f$ is the fermion mass.

We also define

$$N_i=[4(q_i\cdot p_i)]^{-1/2},\eqno(B.3)$$

where $p_i$ and $q_i$ are the momentum and the auxiliary momentum
of the $i$--th massless vector boson, respectively.

Adopting for the polarization
vectors of the gauge bosons the choices (A.12) and (A.16), and replacing
any $p\Dir$ in the fermion propagator numerator with

$$p\Dir =\sum_{\lambda=\pm} u(p,\lambda)\bar u(p,\lambda)\mp m,
\eqno(B.4)$$

one can express the Feynman amplitude $T$ for a
generic diagram as

$$T=\alpha C D M,\eqno(B.5)$$

where $\alpha$ indicates the couplings,
$D$ the appropriate combination of boson and/or fermion
propagators functions, $C$ the eventual color
matrix, and $M$ a combination of $Y$ and $Z$ bilinear spinor functions.

The $Y$ functions are defined as

$$Y(p_1,\lambda_1;p_2,\lambda_2;c_R,c_L)=
[\bar u(p_1,\lambda_1) (c_R P_R + c_L P_L) u(p_2,\lambda_2)].\eqno(B.6)$$

Using (A.4), (A.5) and (A.9) and computing the resulting traces one easily
finds

$$Y(p_1,+;p_2,+;c_R,c_L)=
m_1c_R{{\eta_2}\over{\eta_1}}+m_2c_L{{\eta_1}\over{\eta_2}},\eqno(B.7)$$

$$Y(p_1,+;p_2,-;c_R,c_L)=c_L S(+,p_1,p_2).\eqno(B.8)$$

The remaining $Y$ functions can be obtained by exchanging
$+\leftrightarrow -$ and $R\leftrightarrow L$.

\bigskip
For all processes we will only report those spinor functions $M$
which are not related by a trivial relabeling of momenta and
helicities.
We also adopt the symbol ${\{\lambda\}}$ to denote a set
of helicities of all external particles in a given reaction
and $\sum_{\{\lambda\}}$ to indicate the usual sum over all
possible helicity combinations. The expressions for the couplings
$c_R$ and $c_L$ are given in table III.
The amplitudes squared include the $1/n!$
factor for each $n$--uple of identical final state particles.
Therefore the phase space integration must cover the whole space.
\vskip 0.5in
\noindent
{\bf I. Four fermion production: $\underline{e^-e^+\rightarrow
f\bar f f'\bar f'}$.}
\vskip.25in
The electroweak production of two leptons and two quarks
and of four leptons ($\ell \neq e$)
$$e^-(p_1,\lambda_1)+e^+(p_2,\lambda_2)
\rightarrow
{\ell}^-(p_3,\lambda_3)+{\ell}^+(p_4,\lambda_4)+
f(p_5,\lambda_5)+\bar f(p_6,\lambda_6),\eqno(B.9)$$
is described in the case $f=\ell$ ($f\neq\ell$)
by the eight (first four) Feynman diagrams of fig.12a
plus the four (first two) of fig.12b.
The matrix element is given by

$${\left|{\overline M}\right|}_{{\ell}^-{\ell}^+ f\bar f}^2=
  {{C^f}\over 4} \sum_{\{\lambda\}}
\sum_{l,m=1}^{6}{T}_l^{\{\lambda\}} T_m^{\{\lambda\}*}
\quad\quad(f\neq\ell),\eqno(B.10)$$

$${\left|{\overline M}\right|}_{{\ell}^-{\ell}^+{\ell}^-{\ell}^+}^2=
  {{C^f}\over 16} \sum_{\{\lambda\}}
\sum_{l,m=1}^{12}{T}_l^{\{\lambda\}} T_m^{\{\lambda\}*},
\qquad\qquad\eqno(B.11)$$

with the amplitudes
$$i{T}_{k}=e^4 \sum_{V=\gamma,Z}\sum_{V'=\gamma,Z}
D_k ({M}_{k,VV'}-{1\over{M_Z^2}}N_{k,VV'})
\quad\quad k=1,...6(12),\eqno(B.12)$$
where $D_k$ denotes the propagator functions of the $k$--th diagram.

In formula (B.12) the spinor functions $N_{k,VV'}$ correspond to the
$p^\mu p^\nu$ term of the $Z$--propagator with
momentum $p$, which is zero when acting on a massless fermion
line. By definition $N_{k,V\gamma '}=0$.
For diagrams 5 through 8, 11 and 12 an additional minus sign
has to be inserted because their are related to the
remaining diagrams by the exchange of two identical fermions.

We have

$$ M_{1,{VV'}}=\sum_{i=3,5,6}\sum_{\lambda=\pm}
  Z(p_5,\lambda_5;p_6,-\lambda_6;
    p_3,\lambda_3;p_i,\lambda;
    c_{R_{V'}}^{f},c_{L_{V'}}^{f};
    c_{R_{V'}}^{\ell},c_{L_{V'}}^{\ell})$$
$$\hskip1.2in
\times Z(p_i,\lambda;p_4,-\lambda_4;
         p_2,-\lambda_2;p_1,\lambda_1;
         c_{R_{V}}^{\ell},c_{L_{V}}^{\ell};
         c_{R_{V}}^{e},c_{L_{V}}^{e}),\eqno(B.13)$$

$$\hskip -3.1in
N_{1,VV'}=\sum_{i=3,5,6}\sum_{\lambda=\pm}$$
$$\hskip -.4in
[\sum_{j=5,6}\sum_{\lambda'=\pm}
Y(p_5,\lambda_5;p_j,\lambda';1,1)Y(p_j,\lambda';p_6,\lambda_6,
c_{R_{V'}}^{f},c_{L_{V'}}^{f})]$$
$$\hskip .3in
\times
\{\sum_{k=5,6}\sum_{\lambda''=\pm}
[Y(p_3,\lambda_3;p_k,\lambda'';1,1)Y(p_k,\lambda'';p_i,\lambda;
c_{R_{V'}}^{\ell},c_{L_{V'}}^{\ell})]$$
$$\hskip1.in
\times
Z(p_i,\lambda;p_4,-\lambda_4;
  p_2,-\lambda_2;p_1,\lambda_1;
  c_{R_{V}}^{\ell},c_{L_{V}}^{\ell};
  c_{R_{V}}^{e},c_{L_{V}}^{e})]\},\eqno(B.14)$$

$$\hskip-1.2in
{M}_{2,{VV'}}=-[{M}_{1,{VV'}}(p_1 \leftrightarrow p_2,
p_3 \leftrightarrow p_4,p_5 \leftrightarrow p_6;$$
$$\hskip1.9in
\lambda_1 \leftrightarrow -\lambda_2,\lambda_3\leftrightarrow -\lambda_4,
\lambda_5 \leftrightarrow -\lambda_6)]^{*},\eqno(B.15)$$

$$\hskip-1.2in
{N}_{2,{VV'}}=-[{N}_{1,{VV'}}(p_1 \leftrightarrow p_2,
p_3 \leftrightarrow p_4,p_5 \leftrightarrow p_6;$$
$$\hskip1.9in
\lambda_1 \leftrightarrow -\lambda_2,\lambda_3\leftrightarrow -\lambda_4,
\lambda_5 \leftrightarrow -\lambda_6)]^{*}.\eqno(B.16)$$

\vskip.15in

Then, the spinor functions corresponding to the diagrams of fig.12b are

$$\hskip-2.5in
M_{5(9),{VV'}}=-\sum_{i=1,3,4}b_i\sum_{\lambda=\pm}$$
$$\hskip-.2in
\times Z(p_5,\lambda_5;p_6,-\lambda_6;
    p_2,-\lambda_2;p_i,\lambda;
    c_{R_{V'}}^{f},c_{L_{V'}}^{f};
    c_{R_{V'}}^{e},c_{L_{V'}}^{e})$$
$$\hskip1.2in
\times Z(p_3,\lambda_3;p_4,-\lambda_4;
         p_i,\lambda;p_1,\lambda_1;
         c_{R_{V}}^{\ell},c_{L_{V}}^{\ell};
         c_{R_{V}}^{e},c_{L_{V}}^{e}),\eqno(B.17)$$
where
$$b_1=-b_3=-b_4=-b_5=-b_6=-1.\eqno(B.18)$$

Notice that
$${N}_{k+4(k+8),{VV'}}=0\quad\quad k=1,2(1,...4),\eqno(B.19)$$

since the $Z$--propagators are always connected
to a massless electron line.

Finally, in formulae (B.10)--(B.11) $C^f$ indicates the color factor
$$C^{\ell}=1,\quad\quad{\rm or}\quad\quad C^q=3.\eqno(B.20)$$

\vskip .15 in

The tree--level Feynman diagrams describing the reaction
$$e^-(p_1,\lambda_1)+e^+(p_2,\lambda_2)
\rightarrow
q(p_3,\lambda_3)+\bar q(p_4,\lambda_4)+
q'(p_5,\lambda_5)+\bar q'(p_6,\lambda_6)\eqno(B.21)$$
\vskip.125in
are shown in fig.12a (Only the first four contribute if
$q\neq q'$).
The general form of the amplitude squared is

$${\left|{\overline M}\right|}_{q\bar q q'\bar q'}^2=
  {1\over 4} \sum_{\{\lambda\}} \sum_{l,m=1}^{4}{T}^{\{\lambda\}}_l
T_m^{\{\lambda\}*} C_{lm}
\quad\quad(q\neq q'),\eqno(B.22)$$

$${\left|{\overline M}\right|}_{q\bar q q\bar q}^2=
  {1\over 16} \sum_{\{\lambda\}} \sum_{l,m=1}^{8}{T}^{\{\lambda\}}_l
T_m^{\{\lambda\}*} C_{lm},
\qquad\qquad\eqno(B.23)$$

where $C$ is a $8\times8$ matrix containing the color factors, and

$$iT_k=g_s^2 e^2 \sum_{V=\gamma,Z} D_k M_{k,V}
\quad\quad k=1,...4(8),\eqno(B.24)$$

are the individual amplitudes.

The spinor functions and the propagators can be obtained from
the corresponding expressions
for the case $e^-e^+ \rightarrow {\ell}\bar{\ell} f\bar f$
identifying $V'\rightarrow g$, ${\ell} \rightarrow q$ and $f \rightarrow q'$.

Finally, the matrix of the color factors is

$$C=\left(\matrix{
\alpha & \alpha & \alpha & \alpha & \beta & \beta & \beta & \beta \cr
\alpha & \alpha & \alpha & \alpha & \beta & \beta & \beta & \beta \cr
\alpha & \alpha & \alpha & \alpha & \beta & \beta & \beta & \beta \cr
\alpha & \alpha & \alpha & \alpha & \beta & \beta & \beta & \beta \cr
\beta & \beta & \beta & \beta & \alpha & \alpha & \alpha & \alpha \cr
\beta & \beta & \beta & \beta & \alpha & \alpha & \alpha & \alpha \cr
\beta & \beta & \beta & \beta & \alpha & \alpha & \alpha & \alpha \cr
\beta & \beta & \beta & \beta & \alpha & \alpha & \alpha & \alpha
\cr}
\right),\alpha=2,\beta=-{2\over 3}.\eqno(B.25)$$

\vskip.5in
\noindent
{\bf II. Two fermion and one boson production:
$\underline{e^-e^+\rightarrow f\bar f V}$.}
\vskip.25in

For $Z,\gamma$ production in association with a $f\bar f$ pair ($f \neq e$) in
$$e^-(p_1,\lambda_1)+e^+(p_2,\lambda_2)
\rightarrow
f(p_3,\lambda_3)+\bar f(p_4,\lambda_4)+
Z,\gamma(p_5,\lambda_5),\eqno(B.26)$$
\vskip.11in
the Feynman graphs are depicted in fig.13a and 13b.
For $\gamma$ production the amplitude squared is

$${\left|{\overline M}\right|}_{f\bar f \gamma}^2=
  {{C^f} \over 4} \sum_{\{\lambda\}}
\sum_{l,m=1}^{4}{T}_l^{\{\lambda\}}  T_m^{\{\lambda\}*} ,\eqno(B.27)$$

where

$$-i{T}_{k}={e^3} \sum_{V=\gamma,Z}
D_{k} {M}_{k,V}
\quad\quad k=1,...4.\eqno(B.28)$$

Recalling that $q_i$ is the auxiliary quadrimomentum of the polarization
vector corresponding to the massless boson of momentum $p_i$,
the $M_{k,V}$ functions are

$$\hskip-.8in
{M}_{1,{V}}=N_5 \sum_{i=3, 5}\sum_{\lambda=\pm}
      Z(p_5,\lambda_5;q_5,\lambda_5;
        p_3,\lambda_3;p_i,\lambda;
        1,1;c_{R_{\gamma}}^{f},c_{L_{\gamma}}^{f})$$
$$\hskip 1.in
\times Z(p_i,\lambda;p_4,-\lambda_4;
         p_2,-\lambda_2;p_1,\lambda_1;
         c_{R_{V}}^{f},c_{L_{V}}^{f};
         c_{R_{V}}^{e},c_{L_{V}}^{e}),\eqno(B.29)$$

$${M}_{2,{V}}=-[{M}_{1,{V}}(p_1 \leftrightarrow p_2,
p_3 \leftrightarrow p_4,p_5 \leftrightarrow q_5;
\lambda_1 \leftrightarrow -\lambda_2,
\lambda_3\leftrightarrow -\lambda_4)]^{*},\eqno(B.30)$$

$$\hskip-.4in
{M}_{3,{V}}=N_5 \sum_{i=2,5}b_i\sum_{\lambda=\pm}
      Z(p_3,\lambda_3;p_4,-\lambda_4;
        p_i,\lambda;p_1,\lambda_1;
        c_{R_{V}}^{f},c_{L_{V}}^{f};
        c_{R_{V}}^{e},c_{L_{V}}^{e})$$
$$\hskip 1.2in
\times Z(p_5,\lambda_5;q_5,\lambda_5;p_2,-\lambda_2;p_i,\lambda;
         1,1;c_{R_{\gamma}}^{e},c_{L_{\gamma}}^{e}),\eqno(B.31)$$

$${M}_{4,{V}}=-[{M}_{3,{V}}(p_1 \leftrightarrow p_2,
p_3 \leftrightarrow p_4,p_5 \leftrightarrow q_5;
\lambda_1 \leftrightarrow -\lambda_2,
\lambda_3\leftrightarrow -\lambda_4;
b_1\leftrightarrow b_2)]^{*}.\eqno(B.32)$$

\vskip .25 in

For $Z$ production the amplitude squared has the form

$${\left|{\overline M}\right|}_{f\bar f Z}^2=
  {{C^f}\over 4} \sum_{\{\lambda\}}
{3\over{8\pi m_Z^2}}\int d\Omega\sum_{l,m=1}^{4}{T}_l^{\{\lambda\}}
T_m^{\{\lambda\}*} ,\eqno(B.33)$$

where

$$-i{T}_{k}={e^3\over{sin\theta_W cos\theta_W}} \sum_{V=\gamma,Z}
D_{k} {M}_{k,V}
\quad\quad k=1,...4.\eqno(B.34)$$

The $M_{k,V}$ functions can be obtained from eqs. (B.29) through
(B.32) with the following substitutions: $p_5 \rightarrow r_1$,
$q_5 \rightarrow r_2$, $\lambda_5 \rightarrow -$, $N_5 \rightarrow 1$,
$c_{{L}_{\gamma}}^{e,f} \rightarrow c_{{L}_{Z}}^{e,f}$ and
$c_{{R}_{\gamma}}^{e,f} \rightarrow c_{{R}_{Z}}^{e,f}$
where $r_1$ and $r_2$ are the auxiliary momenta of the
polarization vector for the $Z$.

The color factor is

$$C^{\ell}=1,\quad\quad{\rm or}\quad\quad C^q=3.\eqno(B.35)$$

\vskip .25 in


The Feynman graphs for
$$e^-(p_1,\lambda_1)+e^+(p_2,\lambda_2)
\rightarrow
q(p_3,\lambda_3)+\bar q(p_4,\lambda_4)+
g(p_5,\lambda_5)\eqno(B.36)$$
\vskip.11in
are depicted in fig.13a.
The amplitude squared is

$${\left|{\overline M}\right|}_{q\bar q g}^2=
  {C\over 4} \sum_{\{\lambda\}}
\sum_{l,m=1}^{2}{T}_l^{\{\lambda\}}
T_m^{\{\lambda\}*} ,\eqno(B.37)$$

where

$$-i{T}_{k}=g_s e^2 \sum_{V=\gamma,Z} D_{k} {M}_{k,V}
\quad\quad k=1,2.\eqno(B.38)$$

The spinor functions can be obtained from the corresponding expressions
for $e^-e^+\rightarrow q\bar q \gamma$ substituting
$c_{R_{\gamma}}^{q},c_{L_{\gamma}}^{q}$ with $c_{R_{g}}^{q},c_{L_{g}}^{q}$.

The color factor is
$$C=4.\eqno(B.39)$$

\vskip 0.5in
\noindent
{\bf III. Two fermion and two boson production:
$\underline{e^-e^+\rightarrow f\bar f VV'}$.}
\vskip 0.25in

The Feynman diagrams corresponding to
$$e^-(p_1,\lambda_1)+e^+(p_2,\lambda_2)
\rightarrow
q(p_3,\lambda_3)+\bar q(p_4,\lambda_4)+
g(p_5,\lambda_5)+ g(p_6,\lambda_6)\eqno(B.40)$$

are shown in fig.14a and 14b. The amplitude squared for this process is

$${\left|{\overline M}\right|}_{q\bar q gg}^2=
  {1\over 8} \sum_{\{\lambda\}}
\sum_{l,m=1}^{8}{T}_l^{\{\lambda\}}
T_m^{\{\lambda\}*} C_{lm},\eqno(B.41)$$

with the amplitudes

$$-i{T}_{k}=g_s^2 e^2 \sum_{V=\gamma,Z} D_{k} {M}_{k,{V}}
\quad\quad k=1,...8.\eqno(B.42)$$

Recalling that $q_i$ is the auxiliary quadrimomentum of the polarization
vector corresponding to the massless boson of momentum $p_i$,
one gets for the spinor functions:

$${M}_{1,{V}}=N_5 N_6 \sum_{i=3,5}\sum_{j=4,6}
\sum_{\lambda=\pm}\sum_{\lambda'=\pm}
Z(p_5,\lambda_5;q_5,\lambda_5;
         p_3,\lambda_3;p_i,\lambda;
         1,1;c_{R_{g}}^{q},c_{L_{g}}^{q})$$
$$\hskip.45in
\times Z(p_i,\lambda;p_j,\lambda';
         p_2,-\lambda_2;p_1,\lambda_1;
         c_{R_{V}}^{q},c_{L_{V}}^{q};
         c_{R_{V}}^{e},c_{L_{V}}^{e})$$
$$\hskip1.3in\times
Z(p_6,\lambda_6;q_6,\lambda_6;
  p_j,\lambda';p_4,-\lambda_4;
  1,1;c_{R_{g}}^{q},c_{L_{g}}^{q}),\eqno(B.43)$$

$${M}_{3,{V}}=N_5 N_6
  \sum_{i=3,5}\sum_{j=3,5,6}\sum_{\lambda=\pm}\sum_{\lambda'=\pm}
Z(p_5,\lambda_5;q_5,\lambda_5;
         p_3,\lambda_3;p_i,\lambda;
         1,1;c_{R_{g}}^{q},c_{L_{g}}^{q})$$
$$\hskip-.23in
\times Z(p_6,\lambda_6;q_6,\lambda_6;
         p_i,\lambda;p_j,\lambda';
         1,1;c_{R_{g}}^{q},c_{L_{g}}^{q})$$
$$\hskip.5in
\times Z(p_j,\lambda';p_4,-\lambda_4;
         p_2,-\lambda_2;p_1,\lambda_1;
         c_{R_{V}}^{q},c_{L_{V}}^{q};
         c_{R_{V}}^{e},c_{L_{V}}^{e}),
\eqno(B.44)$$

$$\hskip-1.in
{M}_{5,{V}}=[{M}_{3,{V}}(p_1 \leftrightarrow p_2,
p_3 \leftrightarrow p_4,p_5 \leftrightarrow q_6,
q_5 \leftrightarrow p_6;$$
$$\hskip1.9in
\lambda_1 \leftrightarrow -\lambda_2,
\lambda_3 \leftrightarrow -\lambda_4,
\lambda_5 \leftrightarrow \lambda_6,)]^{*},
\eqno(B.45)$$

$${M}_{7,{V}}= N_5 N_6 \sum_{i=3,5,6}\sum_{\lambda=\pm}\sum_{\lambda'=\pm}
\{2Z(p_3,\lambda_3;p_i,\lambda;
p_5,\lambda_5;q_5,\lambda_5;
c_{R_{g}}^{q},c_{L_{g}}^{q};1,1)$$
$$\hskip1.5in
\times Y(p_6,\lambda_6;p_5,\lambda';1,1)
Y(p_5,\lambda';q_6,\lambda_6;1,1)$$
$$\hskip-.25in
+Z(p_5,\lambda_5;q_5,\lambda_5;p_6,\lambda_6;q_6,\lambda_6;
   1,1;1,1)$$
$$\hskip.1in
\times [Y(p_3,\lambda_3;p_6,\lambda';1,1)
Y(p_6,\lambda';p_i,\lambda;1,1)
-Y(p_3,\lambda_3;p_5,\lambda;1,1)
Y(p_5,\lambda;p_i,\lambda;1,1)]$$
$$\hskip0.1in
-2Z(p_6,\lambda_6;q_6,\lambda_6;
    p_3,\lambda_3;p_i,\lambda;
    1,1;c_{R_{g}}^{q},c_{L_{g}}^{q})$$
$$\hskip1.4in\times
Y(p_5,\lambda_5;p_6,\lambda';1,1)
Y(p_6,\lambda';q_5,\lambda_5;1,1)\}$$
$$\hskip0.8in\times
 Z(p_i,\lambda;p_4,-\lambda_4;
   p_2,-\lambda_2;p_1,\lambda_1;
   c_{R_{V}}^{q},c_{L_{V}}^{q};
   c_{R_{V}}^{e},c_{L_{V}}^{e}),\eqno(B.46)$$

$${M}_{8,{V}}=-[{M}_{7,{V}}(p_1 \leftrightarrow p_2,
p_3 \leftrightarrow p_4;\lambda_1 \leftrightarrow -\lambda_2,
\lambda_3 \leftrightarrow -\lambda_4)]^{*}.
\eqno(B.47)$$

\vskip.15in
In this case, the matrix of the color factors is

$$C=\left(\matrix{
\gamma & \delta & \gamma & \delta & \gamma & \delta & \epsilon & \epsilon \cr
\delta & \gamma & \delta & \gamma & \delta & \gamma & -\epsilon & -\epsilon \cr
\gamma & \delta & \gamma & \delta & \gamma & \delta & \epsilon & \epsilon \cr
\delta & \gamma & \delta & \gamma & \delta & \gamma & -\epsilon & -\epsilon \cr
\gamma & \delta & \gamma & \delta & \gamma & \delta & \epsilon & \epsilon \cr
\delta & \gamma & \delta & \gamma & \delta & \gamma & -\epsilon & -\epsilon \cr
\epsilon & -\epsilon & \epsilon & -\epsilon & \epsilon & -\epsilon & \zeta &
\zeta \cr
\epsilon & -\epsilon & \epsilon & -\epsilon & \epsilon & -\epsilon & \zeta &
\zeta
\cr}
\right),
\gamma={16\over 3},\delta=-{2\over 3},\epsilon=6,\zeta=12.\eqno(B.48)$$

\vskip 0.25in
\noindent
The Feynman diagrams corresponding to the process
$$e^-(p_1,\lambda_1)+e^+(p_2,\lambda_2)
\rightarrow
q(p_3,\lambda_3)+\bar q(p_4,\lambda_4)+
g(p_5,\lambda_5)+ \gamma(p_6,\lambda_6)\eqno(B.49)$$

are depicted in fig.14a and 14c. The amplitude squared is

$${\left|{\overline M}\right|}_{q\bar q g \gamma}^2=
  {C\over 4} \sum_{\{\lambda\}}
\sum_{l,m=1}^{10}{T}_l^{\{\lambda\}}
T_m^{\{\lambda\}*} ,\eqno(B.50)$$

with the amplitudes

$$-i{T}_{k}=\sum_{V=\gamma,Z} g_s e^3 D_{k} {M}_{k,{V}}
\quad\quad k=1,...10.\eqno(B.51)$$

The spinor functions ${M}_{1,{V}}$ through ${M}_{6,{V}}$ can be
obtained from the corresponding expressions for
$e^-e^+\rightarrow q\bar q gg$ substituting $c_{R_{g}}^{q},c_{L_{g}}^{q}$
with $c_{R_{\gamma}}^{q},c_{L_{\gamma}}^{q}$ in the $Z$ functions
which have both $p_6$ and $q_6$ as arguments.
\medskip
The remaining spinor functions are

$${M}_{7,{V}}= N_5 N_6 \sum_{i=3,5}\sum_{j=2,6}b_j
\sum_{\lambda=\pm}\sum_{\lambda'=\pm}
Z(p_5,\lambda_5;q_5,\lambda_5;
         p_3,\lambda_3;p_i,\lambda;
         1,1;c_{R_{g}}^{q},c_{L_{g}}^{q})$$
$$\hskip.45in
\times Z(p_i,\lambda;p_4,-\lambda_4;
         p_j,\lambda';p_1,\lambda_1;
         c_{R_{V}}^{q},c_{L_{V}}^{q};
         c_{R_{V}}^{e},c_{L_{V}}^{e})$$
$$\hskip1.3in\times
Z(p_6,\lambda_6;q_6,\lambda_6;
  p_2,-\lambda_2;p_j,\lambda';
  1,1;c_{R_{\gamma}}^{e},c_{L_{\gamma}}^{e}),\eqno(B.52)$$

$${M}_{8,{V}}=- N_5 N_6 \sum_{i=3,5}\sum_{j=1,6}b_j
\sum_{\lambda=\pm}\sum_{\lambda'=\pm}
Z(p_5,\lambda_5;q_5,\lambda_5;
         p_3,\lambda_3;p_i,\lambda;
         1,1;c_{R_{g}}^{q},c_{L_{g}}^{q})$$
$$\hskip.45in
\times Z(p_i,\lambda;p_4,-\lambda_4;
         p_2,-\lambda_2;p_j,\lambda';
         c_{R_{V}}^{q},c_{L_{V}}^{q};
         c_{R_{V}}^{e},c_{L_{V}}^{e})$$
$$\hskip1.3in\times
Z(p_6,\lambda_6;q_6,\lambda_6;
  p_j,\lambda';p_1,\lambda_1;
  1,1;c_{R_{\gamma}}^{e},c_{L_{\gamma}}^{e}),\eqno(B.53)$$

$${M}_{9,{V}}=[{M}_{8,{V}}(p_1\leftrightarrow p_2,
p_3\leftrightarrow p_4,
\lambda_1 \leftrightarrow -\lambda_2,
\lambda_3 \leftrightarrow -\lambda_4;
b_1\leftrightarrow b_2)]^{*},
\eqno(B.54)$$

$${M}_{10,{V}}=[{M}_{7,{V}}(p_1\leftrightarrow p_2,
p_3\leftrightarrow p_4,
\lambda_1 \leftrightarrow -\lambda_2,
\lambda_3 \leftrightarrow -\lambda_4;
b_1\leftrightarrow b_2)]^{*},
\eqno(B.55)$$

where

$$b_1=b_2=-b_6=-1.\eqno(B.56)$$

Finally, the color factor is
$$C=4.\eqno(B.57)$$

\vfill\eject
\centerline{\bf REFERENCES}
\bigskip
\bigskip
\item{[1]} A. Ballestrero, E. Maina and S. Moretti, Torino Preprint
DFTT 32-92, July 1992. To be published in Phys. Lett. {\bf B}.
\bigskip
\item{[2]} S.L. Wu {\it et al.}, in Proceedings of the ECFA workshop
on LEP 200, A. Bohm and W. Hoogland eds., Aachen FRG, 29 Sept. - 1 Oct. 1986.
CERN 87-08.
\bigskip
\item{[3]} M.L. Mangano and S.J. Parke, Phys. Rep.
{\bf 200} (1991) 301.
\bigskip
\item{[4]} B.L. Ioffe, Phys. Lett. {\bf B78} (1978) 277.
\bigskip
\item{[5]} A. Ali {\it et al.}, Nucl. Phys. {\bf B167} (1980) 454.
\bigskip
\item{[6]} K.J.F. Gaemers and J.A.M. Vermaseren,
Z. Phys. {\bf C 7} (1980) 81.
\bigskip
\item{[7]} D. Danckaert {\it et al.}, Phys. Lett. {\bf B114} (1982) 203.
\bigskip
\item{[8]} R.K. Ellis, D.A. Ross and A.E. Terrano,
Nucl. Phys. {\bf B178} (1981) 421.
\bigskip
\item{[9]} F.A. Berends, P.H. Daverveldt and R. Kleiss,
Phys. Lett. {\bf B148} (1984) 489; Nucl. Phys. {\bf B253} (1985) 441.
\bigskip
\item{[10]} E.W.N. Glover, R. Kleiss and J.J. van der Bij,
Z. Phys. {\bf C 47} (1990) 435.
\bigskip
\item{[11]} E.N. Argyres, C.G. Papadopoulos and S.D.P. Vlassopulos,
Phys. Lett. {\bf B237} (1990) 581.
\bigskip
\item{[12]} R. Kleiss and W.J. Stirling, Nucl. Phys. {\bf B262} (1985)
235.
\bigskip
\item{[13]} C. Mana and M. Martinez,
Nucl. Phys. {\bf B287} (1987) 601.
\bigskip
\item{[14]} K. Hagiwara and D. Zeppenfeld, Nucl. Phys. {\bf B274} (1986)
1.
\bigskip
\item{[15]} G.J. Gounaris, R. Kogerler and H. Neufeld,
Phys. Rev. {\bf D34} (1986) 3257.
\bigskip
\item{[16]} L3 Collaboration, B. Adeva {\it et al.},
Phys. Lett. {\bf B248} (1990) 464.
\item{} OPAL Collaboration, M.Z. Akrawy {\it et al.},
Phys. Lett. {\bf B235} (1990) 389.
\item{} OPAL Collaboration, M.Z. Akrawy {\it et al.},
Z. Phys. {\bf C 49} (1991) 375.
\item{} ALEPH Collaboration, D. Decamp {\it et al.},
Phys. Lett. {\bf B255} (1991) 623.
\item{} DELPHI Collaboration, P. Abreu {\it et al.},
Phys. Lett. {\bf B247} (1990) 167.
\item{} MARK II Collaboration, S. Komamiya {\it et al.},
Phys. Rev. Lett. {\bf 64} (1990) 987.
\item{} DELPHI Collaboration, P. Abreu {\it et al.},
Z. Phys. {\bf C 54} (1992) 55.
\item{} ALEPH Collaboration, D. Decamp {\it et al.},
Phys. Lett. {\bf B284} (1992) 163.
\item{} OPAL Collaboration, P.D. Acton {\it et al.},
Z. Phys. {\bf C 55} (1992) 1.
\bigskip
\item{[17]} L3 Collaboration, B. Adeva {\it et al.},
Phys. Lett. {\bf B271} (1991) 461.
\bigskip
\item{[18]} L3 Collaboration, B. Adeva {\it et al.},
Phys. Lett. {\bf B263} (1991) 551.
\item{} DELPHI Collaboration, P. Abreu {\it et al.},
Phys. Lett. {\bf B274} (1992) 498.
\item{} OPAL Collaboration, G. Alexander {\it et al.},
Z. Phys. {\bf C 52} (1991) 543.
\bigskip
\item{[19]} L3 Collaboration, B. Adeva {\it et al.},
Phys. Lett. {\bf B248} (1990) 227.
\item{} OPAL Collaboration, M.Z. Akrawy {\it et al.},
Z. Phys. {\bf C 49} (1991) 49.
\bigskip
\item{[20]} DELPHI Collaboration, P. Abreu {\it et al.},
Phys. Lett. {\bf B255} (1991) 466.
\item{} ALEPH Collaboration, D. Decamp {\it et al.},
Phys. Lett. {\bf B284} (1992) 151.
\bigskip
\item{[21]} OPAL Collaboration, G. Alexander {\it et al.},
Phys. Lett. {\bf B265} (1991) 462.
\bigskip
\item{[22]} JADE Collaboration, W. Bartel {\it et al.}, Z. Phys.
{\bf C 33} (1986) 23.
\item{} JADE Collaboration, S. Bethke {\it et al.}, Phys. Lett.
{\bf B213} (1988) 235.
\bigskip
\item{[23]} N. Brown and W.J. Stirling, Phys. Lett. {\bf B252} (1990) 657;
Z. Phys. {\bf C 53} (1992) 629.
\bigskip
\item{[24]} ALEPH Collaboration, D. Decamp {\it et al.},
Phys. Lett. {\bf B263} (1991) 112.
\bigskip
\item{[25]} OPAL Collaboration, P.D. Acton {\it et al.},
Preprint CERN-PPE-92-57, Apr. 1992.
\bigskip
\item{[26]} J. Hilgart, R. Kleiss and F. Le Diberder,
Preprint CERN-PPE-92-115, Jul. 1992.
\bigskip
\item{[27]} G.P. Lepage, Jour. Comp. Phys.
{\bf 27} (1978) 192.
\bigskip
\item{[28]} Proceedings of the Workshop on Photon Radiation from Quarks,
S. Cartwright ed., Annecy, France, 2-3 Dec. 1991. CERN 92-04.
\bigskip
\item{[29]} OPAL Collaboration, G. Alexander {\it et al.},
Phys. Lett. {\bf B264} (1991) 219.
\item{} OPAL Collaboration, P.D. Acton {\it et al.},
Z. Phys. {\bf C 54} (1992) 193.
\item{} ALEPH Collaboration, D. Decamp {\it et al.},
Phys. Lett. {\bf B264} (1991) 476.
\item{} DELPHI Collaboration, P. Abreu {\it et al.},
Z. Phys. {\bf C 53} (1992) 555.
\item{} L3 Collaboration, O. Adriani {\it et al.},
Phys. Lett. {\bf B292} (1992) 472.
\par\vfill\eject
\centerline{\bf TABLE CAPTIONS}
\bigskip
\item{table I.} Total cross sections for
$e^+e^-\rightarrow \tau^+ \tau^- \tau^+ \tau^-$,
$e^+e^-\rightarrow \tau^+ \tau^- c \bar c$ and
$e^+e^-\rightarrow \tau^+ \tau^- b \bar b$ at
$\sqrt{s} = 91.1$ and 200 GeV.
Errors are as given by VEGAS [27].
\bigskip
\item{table II.} Total cross sections for
$e^+e^-\rightarrow Z \mu^+ \mu^- $,
$e^+e^-\rightarrow Z \tau^+ \tau^- $ ,
$e^+e^-\rightarrow Z c \bar c$ and
$e^+e^-\rightarrow Z b \bar b$ at
$\sqrt{s} = 200$ GeV.
Errors are as given by VEGAS [27].
\bigskip
\item{table III.} Right and left handed couplings
of the fermions $f=\ell,q$ to the gauge bosons $\gamma,Z,g$.
We have $(e^f, T^f_3, g^f)=(-1, -{1\over 2}, 0)$ for $f=e,\mu,\tau$;
$(e^f, T^f_3, g^f)=(-{1\over 3}, -{1\over 2}, 1)$ for $f=d,s,b$ and
$(e^f, T^f_3, g^f)=({2\over 3}, {1\over 2}, 1)$ for $f=u,c$.

\vfill\eject

$$\vbox{\tabskip=0pt \offinterlineskip
\halign to330pt{\strut#& \vrule#\tabskip=1.em plus2.0em& \hfil#&
\vrule#& \hfil#&
\vrule#& \hfil#&
\vrule#\tabskip=0pt\cr \noalign{\hrule}
&  && && \omit && \cr
&  && &$\hskip3.15cm\sigma_{tot}$& \omit &(pb)\hskip3.25cm& \cr
&  && && \omit && \cr
\noalign{\hrule}
&  && && && \cr
&  &channel\hskip0.6cm& &LEP I ($\sqrt s=91.1$ GeV)& &LEP II
($\sqrt s=200$ GeV)& \cr
&  && && && \cr
\noalign{\hrule}
&  && && && \cr
&  &$\tau^+\tau^-\tau^+\tau^-$& &$(876.4\pm4.3)\times10^{-4}
\hskip0.3cm$& &$(57.8\pm2.0)\times10^{-4}\hskip0.35cm$& \cr
&  && && && \cr
&  &$\tau^+\tau^- c \bar c\hskip0.55cm$& &$(2266.7\pm8.0)\times10^{-4}
\hskip0.2cm$&
&$(269.5\pm3.5)\times10^{-4}\hskip0.25cm$& \cr
&  && && && \cr
&  &$\tau^+\tau^- b \bar b\hskip0.55cm$& &$(406.2\pm1.7)\times10^{-4}
\hskip0.35cm$&
&$(289.9\pm4.7)\times10^{-4}\hskip0.25cm$& \cr
&  && && && \cr
\noalign{\hrule}
\noalign{\smallskip}\cr}}$$

\centerline{Table I}

\vskip 1 in


$$\vbox{\tabskip=0pt \offinterlineskip
\halign to200pt{\strut#& \vrule#\tabskip=1.em plus2.0em& \hfil#&
\vrule#& \hfil#&
\vrule#\tabskip=0pt\cr \noalign{\hrule}
&  && && \cr
&  && &$\sigma_{tot}$ (pb)\hskip1.2cm& \cr
&  && && \cr
\noalign{\hrule}
&  && && \cr
&  &channel\hskip0.1cm& &LEP II ($\sqrt s=200$ GeV)& \cr
&  && && \cr
\noalign{\hrule}
&  && && \cr
&  &$Z^0\mu^+\mu^-$& &$(341.1\pm1.9)\times10^{-3}$\hskip0.25cm& \cr
&  && && \cr
&  &$Z^0\tau^+\tau^-$& &$(170.01\pm0.64)\times10^{-3}$\hskip0.05cm& \cr
&  && && \cr
&  &$Z^0 c \bar c$\hskip0.55cm& &$(402.4\pm1.0)\times10^{-3}$\hskip0.25cm& \cr
&  && && \cr
&  &$Z^0 b \bar b$\hskip0.55cm& &$(378.19\pm0.71)\times10^{-3}\hskip0.05cm$&\cr
&  && && \cr
\noalign{\hrule}
\noalign{\smallskip}\cr}}$$

\centerline{Table II}

\vskip 1in
$$\vbox{\tabskip=0pt \offinterlineskip
\halign to200pt{\strut#& \vrule#\tabskip=1.em plus2.0em& \hfil#&
\vrule#& \hfil#&
\vrule#& \hfil#&
\vrule#& \hfil#&
\vrule#\tabskip=0pt\cr \noalign{\hrule}
&  && && && && \cr
&  && &$~\gamma~~$& &$Z\hskip 1.1truecm$& &$~g~~$& \cr
&  && && && && \cr
\noalign{\hrule}
&  && && && && \cr
&  &$c_R^{f}$& &$e^{f}~$& &$-e^f\sin^2\theta_W\quad$& &$g^f$& \cr
&  && && && && \cr
&  &$c_L^{f}$& &$e^{f}~$& &$T^f_3-e^f\sin^2\theta_W$& &$g^f$& \cr
&  && && && && \cr
\noalign{\hrule}
\noalign{\smallskip}\cr}}$$

\centerline{Table III}

\vfill\eject
\centerline{\bf FIGURE CAPTIONS}
\bigskip
\item{fig.1.} Cross sections for $e^+e^-\rightarrow b \bar b g$
(continuous), $e^+e^-\rightarrow d \bar d g$ (dashed),
$e^+e^-\rightarrow b \bar b gg$ (chain-dotted) and
$e^+e^-\rightarrow d \bar d gg$ (dotted) as a function of $y_{\rm cut}$
for both definitions of $y$ at $\sqrt{s} = 91.1$ GeV.
\bigskip
\item{fig.2.} Cross sections for $e^+e^-\rightarrow b \bar b g$
(continuous), $e^+e^-\rightarrow d \bar d g$ (dashed),
$e^+e^-\rightarrow b \bar b gg$ (chain-dotted) and
$e^+e^-\rightarrow d \bar d gg$ (dotted) as a function of $y_{\rm cut}$
for both definitions of $y$ at $\sqrt{s} = 200$ GeV.
\bigskip
\item{fig.3.} Mass distributions for the $qg$ pair and for the
$q \bar q $ pair in $e^+e^-\rightarrow b \bar b g$
(continuous) and  $e^+e^-\rightarrow d \bar d g$ (dashed)
events in the JADE (3a) and $k_T$ (3b) schemes
at $\sqrt{s} = 91.1$ GeV. All particle pairs have
$y^J_{ij}\geq 1.\times 10^{-2}$ (3a),
$y^T_{ij}\geq 1.5\times 10^{-3}$ (3b).
\bigskip
\item{fig.4.} Cross sections for $e^+e^-\rightarrow q \bar q q \bar q $,
$q = d$ (dotted), $u$ (chain-dotted), $c$ (dashed) and $b$ (continuous)
as a function of $y_{\rm cut}$ for both definitions of $y$
at $\sqrt{s} = 91.1$ GeV.
\bigskip
\item{fig.5.} Cross sections for $e^+e^-\rightarrow q \bar q q' \bar q' $,
$qq' = bd$ (dotted), $du$ (chain-dotted), $bu$ (dashed) and $bc$
(continuous) as a function of $y_{\rm cut}$ for both definitions of $y$
at $\sqrt{s} = 91.1$ GeV.
\bigskip
\item{fig.6.} Two--jet invariant mass distributions in
$e^+e^-\rightarrow b \bar b gg$ at $\sqrt{s} = 91.1$ GeV. The pairs are
$b\bar b$ (continuous), $bg_1$ (dashed), $bg_2$ (chain-dotted) and
$gg$ (dotted) where $g_1$ $(g_2)$ is the most (least) energetic
of the two gluons. All particle pairs have
$y^J_{ij}\geq 1.\times 10^{-2}$ (6a),
$y^T_{ij}\geq 1.5\times 10^{-3}$ (6b).
The angle of each particle with the beam satisfies
$\mid \cos( \theta ) \mid < .95$.
\bigskip
\item{fig.7.} Invariant mass distributions as in fig.6
at $\sqrt{s} = 170$ GeV.
\bigskip
\item{fig.8.} Invariant mass distributions as in fig.6
at $\sqrt{s} = 200$ GeV.
\bigskip\vfill\eject
\item{fig.9.} Invariant mass distributions of
$\tau^+ \tau^-$ pairs in
$e^+e^-\rightarrow \tau^+ \tau^- q \bar q $
at $\sqrt{s} = 170$ GeV, summed over quark
flavours. The invariant mass of the $q \bar q$
pair is required to be greater than 60 GeV in order
to simulate a $Z^0$. The quark masses in this case are neglected.
\bigskip
\item{fig.10.} Cross sections for $e^+e^-\rightarrow b \bar b \gamma$
(continuous), $e^+e^-\rightarrow d \bar d \gamma$ (dashed),
$e^+e^-\rightarrow b \bar b \gamma g$ (chain-dotted) and
$e^+e^-\rightarrow d \bar d \gamma g$ (dotted)
as a function of $y_{\rm cut}$
for both definitions of $y$ at $\sqrt{s} = 91.1$ GeV.
All particle pairs have $y > y_{\rm cut}$.
The photon transverse momentum
is required to be greater than 5 GeV.
\bigskip
\item{fig.11.} Photon energy spectra in
$e^+e^-\rightarrow b \bar b \gamma$ (continuous),
$e^+e^-\rightarrow d \bar d \gamma$ (dashed),
$e^+e^-\rightarrow b \bar b \gamma g$ (chain-dotted)
and $e^+e^-\rightarrow d \bar d \gamma g$ (dotted)
at $\sqrt{s} = 91.1$ GeV.
All particle pairs have
$y^J_{ij}\geq 1.\times 10^{-2}$ (11a),
$y^T_{ij}\geq 1.5\times 10^{-3}$ (11b).
The photon transverse momentum
is required to be greater than 5 GeV.
\bigskip
\item{fig.12.} Feynman diagrams contributing in lowest order to
$e^-e^+\rightarrow q\bar q q'\bar q'$ (12a)
and to $e^-e^+\rightarrow q\bar q \ell \bar \ell$
or $e^-e^+\rightarrow \ell \bar \ell  \ell' \bar \ell'$ (12a and 12b)
with $\ell , \ell' \neq e$.
If the two fermion pairs are not equal only the first four diagrams
in fig.12a and the first two in fig.12b contribute.
A wavy line represents a photon or a $Z^0$ while a jagged line
represents a gluon, a photon or a $Z^0$.
External lines are identified by their momentum as given in the text.
\bigskip
\item{fig.13.} Feynman diagrams contributing in lowest order to
$e^-e^+\rightarrow q\bar q g$ (13a) and to
$e^-e^+\rightarrow q\bar q V$, $V = \gamma,\hskip 2pt Z$ (13a and 13b).
A wavy line represents a photon or a $Z^0$ while a jagged line
represents a gluon, a photon or a $Z^0$.
External lines are identified by their momentum as given in the text.
\bigskip
\item{fig.14.} Feynman diagrams contributing in lowest order to
$e^-e^+\rightarrow q\bar q gg$ (14a and 14b) and to
$e^-e^+\rightarrow q\bar q g\gamma$ (14a and 14c).
A wavy line represents a photon or a $Z^0$ while a jagged line
represents a gluon or a photon.
External lines are identified by their momentum as given in the text.

\par\vfill\eject

\bye